\documentclass[prd,showpacs,preprintnumbers,amsmath,amssymb,floatfix]{revtex4}

\usepackage{graphicx}
\usepackage{dcolumn}
\usepackage{bm}
\usepackage{amssymb}
\usepackage{epsfig}
\usepackage{color}

\newcommand{\ba}{\begin{eqnarray}}
\newcommand{\ea}{\end{eqnarray}}
\newcommand{\be}{\begin{equation}}
\newcommand{\ee}{\end{equation}}
\newcommand{\bdisplay}{\begin{displaymath}}
\newcommand{\edisplay}{\end{displaymath}}

\newcommand{\eq}[1]{Eq.\,(\ref{#1})}

\begin{document}

\title{Implications of a Froissart bound saturation of $\gamma^*$-$p$ deep inelastic scattering. Part II. Ultra-high energy neutrino interactions}
\author{Martin~M.~Block}
\email{mblock@northwestern.edu}
\affiliation{Department of Physics and Astronomy, Northwestern University,
Evanston, IL 60208}
\author{Loyal Durand}
\email{ldurand@hep.wisc.edu}
\altaffiliation{Mailing address: 415 Pearl Ct., Aspen, CO 81611}
\affiliation{Department of Physics, University of Wisconsin, Madison, WI 53706}
\author{Phuoc Ha}
\email{pdha@towson.edu}
\affiliation{Department of Physics, Astronomy and Geosciences, Towson University, Towson, MD 21252}
\author{Douglas W. McKay}
\email{dmckay@ku.edu}
\affiliation{Department of Physics and Astronomy, University of Kansas, Lawrence, KS 66045}



\begin{abstract}
In Part I (in this journal) we argued that  the   structure function $F_2^{\gamma p}(x,Q^2)$ in deep inelastic $ep$ scattering, regarded as a cross section for virtual $\gamma^*p$ scattering, has a saturated Froissart-bounded form  behaving  as $\ln^2 (1/x)$ at small $x$. This form provides an excellent fit to the low $x$ HERA data, including the very low $Q^2$ regions, and can be extrapolated reliably to small $x$ using the natural variable $\ln(1/x)$.  We used  our fit  to derive quark distributions for values of $x$ down to $x=10^{-14}$. We use those distributions here  to evaluate  ultra-high energy (UHE) cross sections for neutrino scattering on an isoscalar nucleon, $N=(n+p)/2$, up to laboratory neutrino energies $E_\nu \sim 10^{16}$-$10^{17}$ GeV where there are now limits on neutrino fluxes.   We estimate that these cross sections are accurate to  $\sim$2\% at the highest energies considered, with the major uncertainty coming from the errors in the parameters that were needed to fit  $F_2^{\gamma p}(x,Q^2)$.   We compare our results to recently published neutrino cross sections derived from NLO parton distribution functions, which become much larger at high energies because of the use of power-law extrapolations of quark distributions to small $x$. We argue that our calculation of the UHE $\nu N$ cross sections is the best one can make based the existing experimental deep inelastic scattering  data. Further, we show that the strong interaction Froissart bound of $\ln^2 (1/x)$ on $F_2^{\gamma p}$ translates to an exact bound of $\ln^3E_\nu$ for leading-order-weak $\nu N$ scattering. The energy dependence of $\nu N$ total cross section measurements consequently has   important implications for hadronic interactions at enormous cms (center-of-mass)  energies not otherwise accessible.
\end{abstract}

\date{\today}

\maketitle


\section{Introduction}
Early in the development of  perturbative QCD (pQCD), the potential for dramatic growth of nucleon structure functions as the Bjorken variable $x$ became small was recognized \cite{GLRsmxqcd}.  Perturbative  analyses showed that the number of low energy gluons rises rapidly as $x$, their fraction of nucleon energy, decreases \cite{desy1990}.  When the collision energy is high enough, large numbers of small $x$ quarks are generated in the parton sea, with interaction energies and momentum transfers large enough to be treated perturbatively. As a result, the  collision cross sections of lepton, photon and hadron collisions on hadrons were predicted to show strong growth at ultra-high energies (UHE), enhancing the prospects for detecting UHE neutrinos of cosmic origin.

This has inspired a number of studies of UHE neutrino cross sections \cite{reno-quigg,fmr,gandhi96,gandhi98}, and  has fundamental implications for the design of experimental cosmic neutrino searches such as the past  searches (AMANDA \cite{amanda}, ANITA \cite{anita1,anita2}, FORTE \cite{forte}, GLUE \cite{glue}, RICE \cite{rice,rice2}), those searches presently underway (ICECUBE \cite{icecube}, Baikal \cite{baikal}, ANTARES \cite{antares}, HiRes \cite{HiRes}, AUGER \cite{auger}), and those experiments under development  (ARA \cite{ara}, ARIANNA \cite{ARI}) or proposed (JEM-EUSO \cite{euso,euso2}). All of these rely on  theoretical models for  total neutrino-nucleon cross sections $\sigma^{\nu(\bar{\nu})}$ at very high energies.  Some searches among those listed have already placed limits on the neutrino flux; others are being designed for discovery of neutrinos above $10^{12}$ GeV \cite{euso,euso2}.  Our work is designed in part with this energy target in mind.

The  proton structure function $F_2^{\gamma p}(x,Q^2)$ for deep inelastic  scattering (DIS) $ep$ scattering has now been measured  at HERA, the electron-proton collider at DESY,  for $x$ in the range $10^{-1}$ to $10^{-6}$,  with virtualities  $Q^2=-\left(e-e'\right)^2$ that ranged from 0.1 GeV$^2$ to 3000 GeV$^2$. The results, as combined  by the H1 and ZEUS detector groups \cite{HERAcombined}, show the expected rapid growth of $F_2^{\gamma p}$, and the quark distributions derived from it, with decreasing $x$ and increasing $Q^2$. However, the smallest values of $x$  attained are still orders of magnitude larger than the values  needed for the reliable calculation neutrino cross sections at UHE.

We argued in  Part I of this paper \cite{bdhmFroissart} that the structure function $F_2^{\gamma p}(x,Q^2)$ is essentially the total cross section for the scattering of an off-shell gauge boson $\gamma^*$ on the proton, a {\em strong interaction process} up to the initial and final gauge boson-quark couplings. In analyzing the HERA data, we therefore chose a fit function that saturated the Froissart bound  \cite{froissart,martin1,martin2} on total hadronic cross sections,  $\sigma(\hat{s} )\propto\ln^2 \hat{s} $ as the  Mandelstam variable $\hat{s}$ for the gauge boson, proton  ($\gamma^* p$) system becomes large,  $\hat{s}\rightarrow\infty$.
This form  is suggested both by theoretical considerations, and by  the remarkably successful descriptions of hadron-hadron and photon-hadron total cross sections over many orders of magnitude obtained with the same saturated functional form in the relevant Mandelstam variables \cite{blockrev}, i.e., $\ln^2 s$. Moreover, the very high energy  proton-proton and proton-air cross sections predicted  using this form for the LHC \cite{LHCtot1,LHCtot2,LHCtot3} and the Pierre Auger Observatory \cite{POAp-air}, respectively, are confirmed by the new data \cite{mbair,blockhalzen,blockhalzen2}. Our high-quality saturated Froissart bound type fit to the small $x$ HERA data \cite{bbmt, bhm,bdhmapp}, extrapolated to ultra-small $x$ using this known functional form as described in Part I,  forms the basis of our present calculations.

We showed in Part I that the quark distributions at small $x$ could be derived from our fit to $F_2^{\gamma p}$ with only the additional input of a relatively small valence-quark contribution $U=u_v\approx 2d_v$ and the small non-singlet combination of quarks, $T_8=u+\bar{u}+d+\bar{d}-2s-2\bar{s}$, which gives the difference between the light- and strange-quark distributions.
Here we apply those quark distributions to a complete calculation of   UHE $\nu N$ cross sections,  in leading order in the weak Fermi coupling).  We include the contributions of the b-quark, omitted in some previous calculations, and the  NLO QCD corrections. We also extend the energy range of earlier calculations up to laboratory neutrino energies $E_\nu= 10^{17}$ GeV, the highest reach of the experimental search for UHE cosmic neutrinos \cite{forte,glue}.

Finally,  we explain how LO UHE neutrino measurements provide an important probe of hadronic interactions at energies far above those that are otherwise accessible, e.g., at average final center-of-mass hadronic energies $W_{av}>900$ TeV for $E_\nu>10^{13}$ GeV.  We  show that our {\em first-order-weak $\nu N$ cross sections are  bounded by $\ln^3 E_\nu$ at UHE}. This is a direct consequence of the effectively hadronic scattering of the off-mass shell gauge boson $W^*$ or $Z^*$ on $N$  being Froissart-bounded by $\ln^2(1/x)$, and provides a test of this picture and of the hadronic Froissart bound more generally.

The organization of the paper is as follows.
In Sec.\ \ref{sigmaderivation} we review the structure of the  $\nu N$ and $\bar{\nu} N$ charged and neutral current differential and total cross sections, and the  quark-parton construction of of the DIS structure functions on which they depend. In  Sec.\ \ref{subsec:predictions}, we present  the total cross sections calculated using the quark distributions derived in Part I from our Froissart-bounded fit to the HERA data on $F_2^{\gamma p}$, as extrapolated to very small $x$. We examine the sensitivity of the results to different regions in $x$ and $Q^2$ and the different structure functions in Sec.\ \ref{subsec:sensitivity}. We compare our results with those of other calculations based on quark distributions obtained in standard perturbative analyses of the HERA data, and then extrapolated to very small $x$, in Sec.\ \ref{subset:comparisons}.

In Section {\ref{sec:hadronic_implications}, we examine the importance of the detection and measurement of the cross section of UHE neutrinos as a {\em new and powerful method of probing   UHE hadronic  physics (QCD)}. We also present analytic expressions for $\sigma _{CC}(E\nu)$ and $\sigma _{NC}(E\nu)$ at high energies $E_\nu$. We summarize and draw conclusions in Sec. \ref{conclusions}, and present some details with respect to the calculations, as well as a comparison with calculations based on the supposed wee parton limit, in two Appendices.


\section{Ultra-high energy $\nu N$ cross sections \label{sigmaderivation}}

\subsection{Differential cross sections}

Expressions for the general quark parton  charged current (CC) and neutral current (NC) $\nu N$ cross sections are given in many references; see for example \cite{c-ss,gandhi98,c-sms,PDG}. We display them here to keep our presentation self-contained, using the notation of Ref. \cite{c-sms}.  The LO  double differential inclusive charged current cross sections for neutrino or antineutrino scattering on an isoscalar nucleon target ${N}=(n+p)/2$,  $\nu_\ell+N\rightarrow \ell+X$ or $\bar{\nu}_\ell+N\rightarrow \bar{\ell}+X$, $\ell=e,\,\mu,\,\tau$, are
\ba
\frac{d^2\sigma^{\nu(\bar{\nu})}_{CC}}{dxdQ^2}(E_{\nu},Q^2,x) &=& \frac{G_F^2}{4\pi}\left(\frac{M_{W}^2}{Q^2+M_{W}^2}\right)^2
 \nonumber \\
 && \times\frac{1}{x}\left[F^{\nu(\bar{\nu})}_2\pm xF^{\nu(\bar{\nu})}_3
+ (F^{\nu(\bar{\nu})}_2\mp xF^{\nu(\bar{\nu})}_3)\left(1-\frac{Q^2}{2mxE_\nu}\right)^2-\left(\frac{Q^2}{2mxE_\nu}\right)^2F^{\nu(\bar{\nu})}_L\right].
\label{masterEq2cc}
\ea
The upper signs are for $\nu$ and the lower, for $\bar\nu$.

The corresponding double differential cross sections for the neutral current processes $\nu_\ell+N\rightarrow \nu_\ell +X$,  $\bar{\nu}_\ell+N\rightarrow\bar{ \nu}_\ell +X$ are
\ba
\frac{d^2\sigma^{\nu(\bar{\nu})}_{NC}}{dxdQ^2}(E_{\nu},Q^2,x) &=& \frac{G_F^2}{4\pi}\left(\frac{M_{Z}^2}{Q^2+M_{Z}^2}\right)^2  \nonumber \\
&& \times\frac{1}{x}\left[F0^{\nu(\bar{\nu})}_2\pm xF0^{\nu(\bar{\nu})}_3
+(F0^{\nu(\bar{\nu})}_2\mp xF0^{\nu(\bar{\nu})}_3)\left(1-\frac{Q^2}{xs}\right)^2-\left(\frac{Q^2}{xs}\right)^2F0^{\nu(\bar{\nu})}_L\right].
\label{masterEq2nc}
\ea
The NC chiral coefficients are defined as $L_u=1-\frac{4}{3}\sin^2\theta_W, L_d= -1+\frac{2}{3}\sin^2\theta_W, R_u=-\frac{4}{3}\sin^2\theta_W$ and $R_d=\frac{2}{3}\sin^2\theta_W$. The value $\sin^2\theta_W$=0.231 \cite{PDG} was used in all the present calculations.

In these expressions, $x=Q^2/2p\cdot q$ is the Bjorken scaling variable, $p$ is the nucleon 4 momentum, and $Q^2=-q^2$ where $q=\nu-\ell$ is the momentum of the virtual $W$ or $Z$ boson  which interacts with the nucleon, that is, the momentum transferred from the leptons in the scattering. The second independent scaling variable is the fraction of the neutrino energy transferred  to the hadronic system, $y=(E_\nu-E_\ell)/E_\nu$ in the nucleon rest frame; clearly, $0\leq y\leq 1$. In covariant form, $y=p\cdot q/p\cdot \nu= Q^2/2mxE_\nu$, as it appears in Eqs.\ (\ref{masterEq2cc}) and (\ref{masterEq2nc}).

The direct channel $\nu N$ Mandelstam variable $s$ is $s=(\nu+p)^2=2 m E_{\nu}$, $m$ the nucleon mass, where we neglect $m^2$ relative to $2mE_\nu$, so $y=Q^2/xs$.  Also, the direct channel Mandelstam variable $\hat{s}$ for the  strong $W^*N$ or $Z^*N$ scattering is $\hat{s}=(q+p)^2=2p\cdot q-Q^2+m^2\approx Q^2/x$. The structure functions $F_1,\,F_2,\,F_3,$ and $F_L=F_2-2xF_1$ are functions of the  Bjorken variable $x$ and the virtuality $Q^2$ of the gauge boson.

The functions $F_i=F_1,\,x^{-1}F_2,\,F_3$ are given in terms of parton distribution functions (PDFs) by expressions   $F_{i,0}$ of the LO form convoluted  with QCD correction terms  \cite{hw,furmanski,esw}, schematically
\be
\label{Fi_PDF_connection}
F_i = \left[\openone+\frac{\alpha_s}{2\pi}C_{iq}\right]\otimes F_{i,0} + \frac{\alpha_s}{2\pi}C_{ig}\otimes g.
\ee
Here  $I$ is the unit operator,  $C_{iq}$ and $C_{ig}$ are coefficient functions from the operator product expansion, known in low order. The symbol $\otimes$ indicates  convolution of $C_{iq},\,C_{ig}$ with $F_{i,0}$ and the gluon distribution $g=g(x,Q^2)=x^{-1}G(x,Q^2)$.

The uncorrected structure functions $F_{i,0}$ are sums of quark distributions
\be
\label{Ff0_defined}
F_{i,0}(x,Q^2) = \sum_j c_{i,j}\,q_j(x,Q^2),
\ee
with the constants $c_{i,j}$ being the weights with which the different quarks appear. In particular, the structure functions $F^{\nu}_{1,0}(x,Q^2),\, F^{\nu}_{2,0}(x,Q^2)$ and $xF_{3,0}^{\nu}(x,Q^2)$ for neutrino scattering on an isoscalar nucleon $N=(p+n)/2$ are expressed in terms of the quark distributions $u=u(x,Q^2),\,d= d(x,Q^2),\,s= s(x,Q^2), \ldots$, as follows:
\ba
\label{F10CC}
F_{1,0}^\nu &=& \frac{1}{2}x^{-1}F_{2,0}^\nu, \label{F1}\\
\label{F20CC}
x^{-1}F_{2,0}^\nu &=& u+d+\bar{u}+\bar{d}+2s+2\bar{c}+2b+\ldots, \label{F2} \\
\label{F30CC}
F_{3,0}^\nu &=& u+d-\bar{u}-\bar{d}+2s-2\bar{c}+2b-\ldots. \label{F3}
\ea
Similarly,
\ba
\label{F10CCnubar}
F_{1,0}^{\bar{\nu}} &=& \frac{1}{2}x^{-1}F_{2,0}^{\bar{\nu}}, \label{F1nubar}\\
\label{F20CCnubar}
x^{-1}F_{2,0}^{\bar{\nu}} &=& u+d+\bar{u}+\bar{d}+2\bar{s}+2c+2\bar{b}+\ldots, \label{F2nubar} \\
\label{F30CCnubar}
F_{3,0}^{\bar{\nu}} &=& u+d-\bar{u}-\bar{d}-2\bar{s}+2c-2\bar{b}+\ldots. \label{F3nubar}
\ea

The corresponding expressions for the neutrino neutral current structure functions and cross sections are
 \ba
 \label{F10NC}
 F0_{1,0}^{\nu(\bar{\nu})} &=& \frac{1}{2}x^{-1}F0^{\nu(\bar{\nu})}_{2,0}, \\
x^{-1}F0^{\nu(\bar{\nu})}_{2,0}&=&\left(d+u+\bar{d}+\bar{u}\right)\left(L_u^2+R_u^2+L_d^2+R_d^2\right)/4 \nonumber \\
 \label{F20NC}
& & + \left(s+b+\bar{s}+\bar{b}\right)\left(L_d^2+R_d^2\right)/2+\left(c+\bar{c}\right)\left(L_u^2+R_u^2\right)/2, \\
\label{F30NC}
F0^{\nu(\bar{\nu})}_{3,0}&=& \pm \left(d+u-\bar{d}-\bar{u}\right)\left(L_u^2-R_u^2+L_d^2-R_d^2\right)/4.
\ea

In these expressions, we have referred all distributions to the proton so that $d_n\rightarrow u_p\equiv u,\ u_n\rightarrow d_p\equiv d$, etc. We will take  $\bar{s}=s,\ \bar{c}=c,\ \bar{b}=b$ since these quarks are produced only in pairs through gluon splitting. It is then clear that $F_{2,0}^{\bar{\nu}}=F_{2,0}^\nu$ and that $F_{3,0}^{\bar{\nu}}$ and $F_{3,0}^\nu$ would have opposite signs except for the presence of the valence distributions $u_v=u-\bar{u}$, $d_v=d-\bar{d}$. Valence-quark effects are unimportant at small $x$, the region of primary interest here, so  $F_{3,0}^{\bar{\nu}}\approx-F_{3,0}^\nu$.

The uncorrected longitudinal structure functions $F_{L,0}=F_{2,0}-2xF_{1,0}$ are equal to zero in leading order in the strong coupling, but the corrected functions $F_L$  gain terms of orders $\alpha_s/(2\pi)$ and higher through the convolutions of the different coefficient functions for $F_1$ and $x^{-1}F_2$ with the quark and gluon PDFs \cite{esw}.


\subsection{Total cross sections at UHE\label{subsec:total_sigmas}}


Integration of Eqs.\ (\ref{masterEq2cc}) and (\ref{masterEq2nc}) over the allowed ranges in $x$ and $Q^2$ gives the total CC and NC $\nu N$ cross sections to {\em  leading order in the weak coupling $G_F$, but all orders in the strong hadronic interactions}. The integrations are bounded by the conditions $x\leq 1$ and $0\leq y\leq1$ with $y=Q^2/2mE_\nu x$. We limit ourselves to a minimum value of $Q^2$ that is consistent with the application of perturbative QCD and at the same time avoids possible problems with numerical instabilities as $x$ or $Q^2$ goes to zero. We therefore take $Q^2$ and $x$ in the ranges $s=2mE_\nu\geq Q^2\geq Q^2_{min} \simeq 1 {\rm \ GeV}^2$ and $Q^2/(2mE_{\nu})\leq x\leq 1$, integrating first over $x$.

As emphasized in early calculations (see  Ref.  \cite{mr85}), the vector boson (V) propagator factor ($M_{V}^2/(M_{V}^2+Q^2))^2$  cuts off the integrand for $Q^2 \gg M_{V}^2$, effectively selecting a range of small $x$ reaching somewhat below $x\sim M_{V}^2/(2mE_{\nu})$ which makes the only substantial contributions to the total cross section.  For the range of neutrino energies we consider in this work,  $10^6$ GeV $\leq E_{\nu}\leq 10^{17}$ GeV, this means that we must accurately probe $x$ values in the range $0.001\leq x\leq10^{-14}$.

With the above discussion in mind, we express the total CC cross sections as
\ba
\sigma^{\nu{\rm N}}_{CC}(E_{\nu})&=& \int_{Q_{min}^2}^sdQ^2\int_{Q^2/s}^1dx\frac{d^2\sigma_{CC}}{dxdQ^2}(E_{\nu},Q^2,x)\nonumber\\
&=&\frac{G_F^2}{4\pi}\int_{Q_{min}^2=1}^{2mE_\nu}dQ^2\left(\frac{M_{W}^2}{Q^2+M_{W}^2}\right)^2\int_{Q^2/(2 m E_\nu)}^1\frac{dx}{x} \nonumber \\
&& \times\left[F^{\nu}_2+xF^{\nu}_3+(F_2^{\nu}-xF_3^{\nu})\left(1-\frac{Q^2}{xs}\right)^2-\left(\frac{Q^2}{xs}\right)^2\!\!F_L^{\nu}\right].
\label{totEq2cc}
\ea
 The corresponding  total NC cross section, $\sigma^{\nu{\rm N}}_{NC}(E_{\nu})$ is obtained by the replacement of $M_W$ by $M_Z$ and the replacement of $F_2^{\nu}$,  $xF_3^{\nu}$ and $F_L^{\nu}$ by $F0_2^{\nu}$, $xF0_3^{\nu}$ and $F0_L^{\nu}$.

Written in this form, the integration over $x$ shows clearly that a simple power law behavior of $F_2^{\nu(\bar{\nu})}$ as a function of $x$ or  $\ln(1/ x)$ will produce the same power of $x$, or an added power of $\ln(1/x)$, in the result.  This is helpful for assessing the UHE behavior of the total cross section that follows from a given model of the structure functions, as pointed out in \cite{kniehl}. In particular, given the effective cutoff  in the $Q^2$ integration for $Q^2>M_W^2$, it shows that when the neutrino energy  $E_\nu$ satisfies the condition $E_\nu/M_W^2\gg 1$, {\em the $\nu N$ cross section calculated to lowest order in $G_F$ will rise asymptotically with neutrino energy  as $ \ln^3E_\nu$ for  our Froissart-bounded extrapolations of $F_2^{\nu(\bar{\nu})}$\,}\footnote{The authors of \cite{kniehl} incorrectly state that the work in \cite{bbmt} and \cite{bhm} claims that the $\nu N$ cross section rises only as $\ln^2{E_{\nu}}$ in the large $E_{\nu}$ limit.  Refs.\ \cite{bbmt} and \cite{bhm} actually assume the Froissart saturated form {\em only} for the $x$ dependence of the structure function $F_2^{\gamma p}$ and {\em not} for the integrated cross sections $\sigma^{\nu(\bar{\nu})}$ This confusion between the gauge boson-$N$ cross section and the total $\nu N$ cross section is clarified in \cite{bhm11}. }.


\section{Predictions for ultra-high energy neutrino cross sections  \label{ultrahighsigma}}

\subsection{Results for ultra-high energy  neutrino cross sections from the Froissart-bounded fit to $F_2^{\gamma p}$ \label{subsec:predictions}}


As shown in detail in Part I, it is straightforward to derive quark distributions at very small $x$ from our Froissart-bounded fit to $F_2^{\gamma p}$ as extrapolated to that region. The extrapolation should be quite reliable: the fit to $F_2^{\gamma p}$ is excellent as detailed in \cite{bdhmapp}, the Froissart-bounded fit function reduces to a simple quadratic in $\ln(1/x)$ for $x$ small, and this need only be extrapolated by a factor of $\sim 2.7$ in $\ln(1/x)$ to get to from the lower end of the HERA region, $x\sim 10^{-5}$, to the smallest values of $x$ needed, $x\sim 10^{-13}$-$10^{-14}$. The expected correlated statistical errors in the result are $\sim 1$-2\%. The QCD corrections in \eq{Fi_PDF_connection}, necessary to obtain the final structure functions from their quark-level expressions, will be treated in NLO. These corrections can be calculated analytically at small $x$, and their calculation does not introduce further errors in the cross sections. Details of the analytic calculations are given in the Appendices to Part I, with some further detail in the present Appendices. Given our analytic expressions for the small-$x$ structure functions, integration of Eqs.\ (\ref{masterEq2cc}) and (\ref{masterEq2nc}) over $x$ and $Q^2$ as in \eq{totEq2cc}  gives the expected neutrino cross sections.

We limit our calculations to five quark flavors, allowing $n_f$ to change with $Q^2$ as in Part I. Because of the effective $Q^2$ cutoff imposed by the gauge-boson propagator factor $1/(Q^2+M_V^2)^2$ in the integrand, the top quark, which becomes active as a parton only for $Q^2\gg m_t^2\approx (173)^2$ GeV$^2$,  does not contribute significantly to the cross sections and we neglect its contribution. The $b$ quark, active for $Q^2>m_b^2$, contributes to the CC cross section only when the threshold condition $\hat{s}\approx Q^2/x>m_t^2$ for the process $W^++b\rightarrow t$  is satisfied. This condition is satisfied over essentially all of the important region in $x,\,Q^2$ space.

In Fig.\ \ref{fig:heraCC-NC} we show the  charged current and neutral current neutrino cross sections for large $E_\nu$ calculated using our complete results. We believe these give the best predictions that can be made using current experimental information. They do not depend on the large  extrapolations  of the starting  PDFs  used in standard analyses---typically with power dependence in $x$---over many orders of magnitude in $1/x$; they depend only on the extension of our   physically motivated Froissart-bounded fit to the HERA data on $F_2^{\gamma p}$ over a factor of $\approx 2.7$ in the natural variable $v=\ln{(1/x)}$.

%
\begin{figure}[htbp] 
\includegraphics{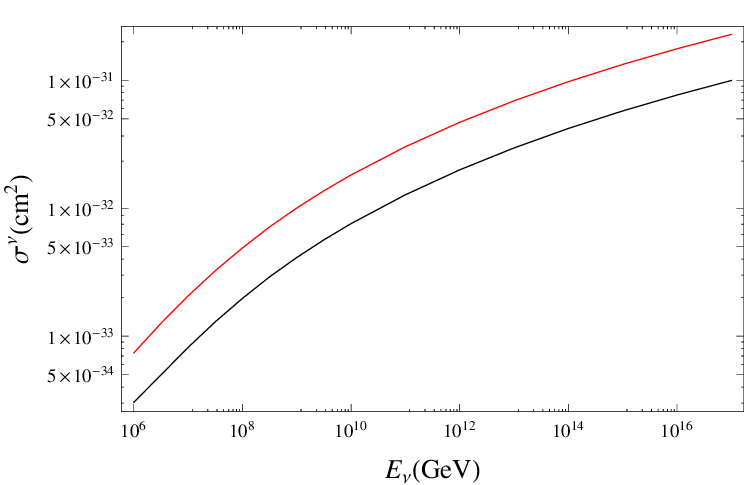}
 \caption{Plots of the $\nu N$ cross sections, in cm$^2$, vs.  $E_\nu$, the laboratory neutrino energy, in GeV,  calculated using the  extrapolation of the global fit to the HERA data on $F_2^{\gamma p}(x,Q^2)$ to small $x$  discussed in Part I \cite{bdhmFroissart} and the relations between $F_2^{\nu(\bar{\nu})}$, $F0_2^{\nu(\bar{\nu})}$, and $F_2^{\gamma p}$ in Eqs.\ (\ref{F2nu_F2_connection_nf=5}) and (\ref{F02nu_F2_connection_smallx}), with  NLO treatments of the small functions $T'_i$ and of the subdominant structure functions $F_3^{\nu(\bar{\nu})}$ and $F_L^{\nu(\bar{\nu})}$.  The upper curve (red) is the CC cross section and lower curve (black) is the NC cross section.}
   \label{fig:heraCC-NC}
\end{figure}

We give numerical values for the cross sections in the first columns in Tables \ref{table:CCx-secs} and \ref{table:NCx-secs}.


\subsection{Sensitivity of the neutrino cross sections to $x$ and $Q^2$ for changing $E_\nu$ \label{subsec:sensitivity}}


As the neutrino energy increases, the $W^* N$ ($Z^*N$) interaction probes more deeply into the small $x$ region of the nucleon.  We illustrate this effect in Fig. \ref{fig:HERAx-dist}, where we show the differential distribution of contributions to the CC cross section with respect to $ x$, normalized to the total cross section.  The dominant contributions to $\sigma^{\nu N}(E_{\nu})$ march steadily downward in $x$ as $E_{\nu}$ increases.  For example, at $E_{\nu} = 10^{12}\ {\rm  GeV}$, the region in $x$ from $10^{-9}$ to $10^{-4}$ determines the value of $\sigma^{\nu N}(E_{\nu})$.
%
 \begin{figure}[htbp]
\includegraphics{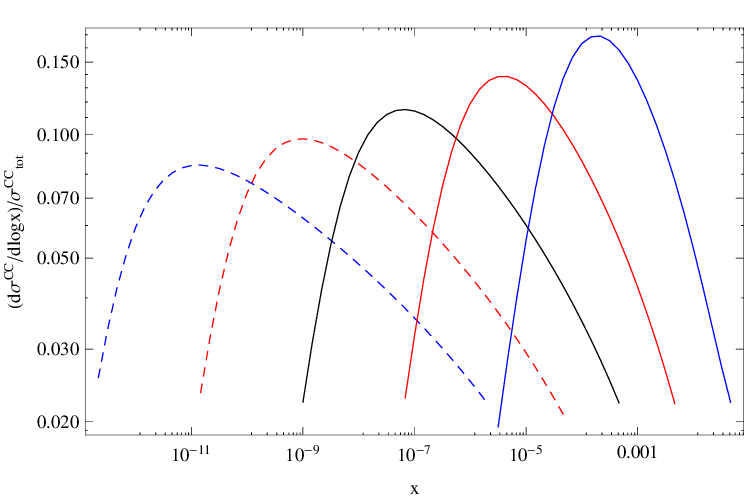}
 \caption{The normalized distributions of contributions to the $\nu N$  total cross CC section of Fig.\ \ref{fig:heraCC-NC} with respect to $x$ for cross sections  at energies $ 10^8, \ 10^{10},\  10^{12},\ 10^{14}$ and $10^{16}$ GeV, right to left. The integration over the variable $x$ involves $dx/x$, so the $\ln x$ distribution is the natural measure.}
   \label{fig:HERAx-dist}
\end{figure}
%

Given this information, we see that our extrapolation of $F_2^{\gamma p}$ into the ultra-small $x$ region is much less extreme than it might be, since the Froissart-like form for $F_2^{\gamma p}(x,Q^2)$ reduces to a quadratic form in  $v=\ln(1/x)$ for $x$ small. The bulk of HERA data are in the range from $v\approx 2.3$ to $v\approx 10$. It is evident from Fig.\ \ref{fig:HERAx-dist} that the most important region for the prediction of the neutrino cross section at $E_\nu=10^{16}$ GeV is around $x=10^{-11}$ or $v=25$, an extrapolation of roughly a factor of 2.5 in this variable. The fact that the small $x$ or large $v$ form of $F_2^{\gamma p}$, a simple quadratic in $v$, is tightly constrained theoretically by the Froissart limit, and experimentally by the quality of the fit, limits the uncertainties introduced by the extrapolation.

The values of the CC and NC cross sections and their fractional uncertainties  calculated from the squared error matrix of the fit (including correlation errors) over the energy range from $10^6$ GeV up to $10^{17}$ GeV are shown in Table \ref{tab:errors}. The integration errors are negligible, a part in $10^6$ or better. As is expected for a linear fit, calculated uncertainties  of 1\% to 2\% are similar is size to the errors in the fit parameters themselves.  As far as purely numerical uncertainties are concerned, these are the smallest one can obtain within the range of current efforts to estimate the UHE CC and NC neutrino-nucleon cross sections.
%
\begin{table}[htbp]
   \begin{center}
   \renewcommand{\arraystretch}{1.5}
   \begin{tabular}{|c||c|c|c|c|c|c|c|c|c|c|c|c|}
         \hline
       Energy  (GeV) &$10^6$&$10^7$&$10^8$&$10^9$&$10^{10}$&$10^{11}$ & $10^{12}$ & $10^{13}$ & $10^{14}$ & $10^{15}$ & $10^{16}$ & $10^{17}$ \\ \hline\hline

        $\sigma_{{\rm CC}}{\rm \ (nb) }$&0.740&2.06&4.89&10.0&18.2&30.2&46.9&69.1&97.5&133&176&228 \\ \hline
        $\delta \sigma_{{\rm CC}}/\sigma_{{\rm CC}}$ &0.009&0.010&0.012&0.014&0.016&0.017&0.017&0.014&0.015&0.015&0.016&0.017 \\ \hline
        $\sigma_{{\rm NC}}{\rm\  (nb})$&0.304&0.817&1.97&4.12&7.58&12.7&20.0&29.6&42.0&57.6&76.7&99.6 \\ \hline
        $\delta \sigma_{{\rm NC}}/\sigma_{{\rm NC}}$ &0.010&0.010&0.013&0.015&0.016&0.017&0.017&0.014&0.015&0.016&0.016&0.017 \\ \hline
   \end{tabular}
   \renewcommand{\arraystretch}{1}
   \end{center}
   \caption{Total cross sections in nb and the correlated fractional errors in the  $\nu N$ CC and NC cross sections,  as a function of neutrino energy in GeV, computed from the global saturated Froissart-bounded fit to the  $F^{\gamma p}_2(x,Q^2)$ HERA data.}
   \label{tab:errors}
\end{table}

The $Q^2$ dependence of the $\nu N$  CC cross section  is shown in Fig.\  \ref{fig:HERAQsq-dist}, where we plot  the partial cross sections obtained by integrating \eq{masterEq2cc} first over $x$, and then over $Q^2$ with $Q^2\geq Q_0^2$, for a selection of values of $Q_0^2$ from 1 GeV$^2$  to $s=2m E_{\nu}$. The partial cross sections are  normalized by the total CC cross section and plotted against $Q^2_0$, the  minimum  value of $Q^2$ included.  The curves show the results for $E_{\nu}$ = $10^6,\ 10^8,\ 10^{10},\ 10^{12},\ 10^{14}$ and $10^{16}$ GeV.

Combining these curves with those in Fig.\ \ref{fig:HERAx-dist}, we can now identify the major contributors to the total cross section in \eq{totEq2cc}.  We note that if $y\equiv  Q^2/(2xmE_\nu)\approx 0$, {\em only} the structure function $F_2^{\nu}$ contributes to the cross section, with a coefficient of 2. As an example, at $E_\nu=10^{12}$ GeV, we see from Fig.\  \ref{fig:HERAx-dist} that the most likely value of $x$ is $\approx 10^{-7}$, and from  Fig.\  \ref{fig:HERAQsq-dist}, that  $\approx 90$\% of the cross section comes from $Q^2$ smaller than about $4\times10^4$ GeV$^2$. Thus, about 90\% of the time, $y<0.2$; so the coefficient of $F_2^\nu >1.6$. A complete calculation shows that the average value of $y$ over the distribution ranges from $\sim 0.2$ at $E_\nu=10^6$ GeV to $\sim0.08$ at $10^{16}$ GeV.  Similarly, we estimate the typical coefficient of $F_3^{\nu(\bar{\nu})}$, itself $<(1/5)F_2^{\nu(\bar{\nu})}$, as $<0.4$, for a contribution relative to that of $F_2^{\gamma p}$ of $\sim 0.05$. The coefficient of the NLO function $F_L^\nu$, whose evaluation requires a knowledge of the gluon distribution in addition to  $F_{2,0}^{\gamma p}$, is $y^2\sim 0.04$.  Thus, we see that the overwhelmingly dominant  contribution to the neutrino cross section is from $F_2^\nu$.

Further, as shown in Part I and discussed in Appendix\ \ref{sec:F2},  $F_2^{\nu(\bar{\nu})}$ is given  to all orders in $\alpha_s$  as a simple multiple of $F_2^{\gamma p}$ up to small additive non-singlet corrections which are known to NLO,  sufficient accuracy for our purposes. The sub-dominant functions  $F_3^{\nu(\bar{\nu})}$ and $F_L^{\nu(\bar{\nu})}$ to $F_2^{\gamma p}$, suppressed in the cross sections by the factors estimated above, are also known to NLO. All are included in our final cross section calculations. The CC $\nu N$ cross section is therefore determined to high accuracy by $F_2^{\gamma p}$, a quantity obtained from the fit to experimental data.  Similar arguments hold for the NC cross sections.

We also see from  Fig.\  \ref{fig:HERAQsq-dist} that at least 90\% of the cross section is captured for a lower cutoff in the $Q^2$ integration of $Q_0^2 = 250$ GeV$^2$ for $E_\nu=10^6$ GeV,  500 GeV$^2$ for $E_\nu=10^8$ GeV,  and $Q_0^2= 1000$ GeV$^2$ for $E_{\nu} > 10^9$ GeV.  The final state hadronic invariant mass $W^2 \sim Q^2/x\gg Q^2$ is therefore greater than $m_t^2$  over all of the important region of $x,\,Q^2$ space in the ultra-high energy CC and NC cross section integrands, e.g., for $x<0.017$ for $Q^2=500$ GeV$^2$. The process $W^++b\rightarrow t$ is therefore allowed and the $b$-quark contribution is essential for the calculation of  the  ultra-high energy CC and NC cross sections. This contribution was omitted in early work.

 \begin{figure}[htbp]
\includegraphics{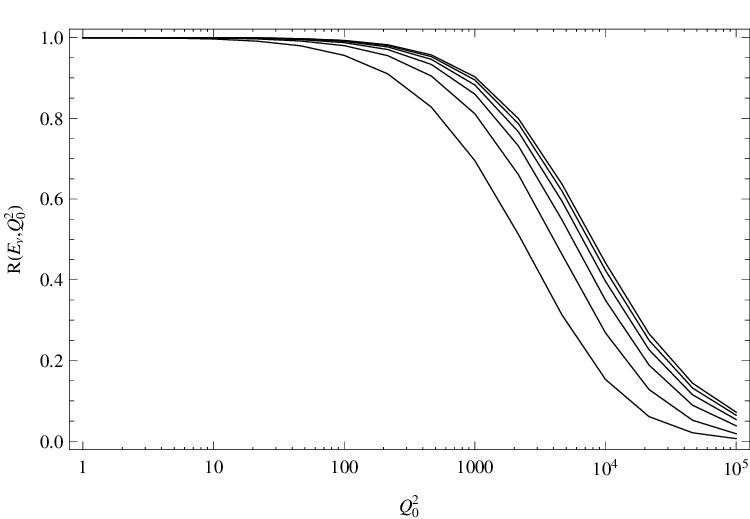}
 \caption{The normalized distributions of contributions to the $\nu N$ total cross CC section of Fig.\ \ref{fig:HERAQsq-dist} with respect to $Q_0^2$, in GeV$^2$, for cross sections  at energies $10^6$, $10^8$, $10^{10}$, $10^{12}$ $10^{14}$ and $10^{16}$ GeV, bottom to top. After integration over $x$, which starts at $Q^2/s$, the integration over the variable $Q^2$ starts at the minimum value $Q^2_0$  consistent with the range of validity of the determination of $F^{\nu}_2(x,Q^2)$. The plot shows that the calculated value of the UHE cross section is very insensitive to $Q^2_0$, the lowest value of the integration range over $Q^2$.}
   \label{fig:HERAQsq-dist}
\end{figure}


\subsection{Comparisons with  predictions based on extrapolated PDFs  \label{subset:comparisons}}


The application of solutions to the DGLAP equations for quark PDFs to evaluation of UHE $\nu N$ CC and NC cross sections has a long history.  The application reported by Ghandi, Quigg, Reno and Sarcovic (GQRS) \cite{gandhi98}, based on $d, u, s$ and $c$ quark PDFs from 1998 CTEQ4 analysis of the early ZEUS small $x$ data, was a standard for many years, and still remains a point of comparison. The recent results of Connolly, Thorne and Waters (CTW) \cite{ctw}, and those of Cooper-Sarkar, Mertsch and Sarkar (CSMS) \cite{c-sms}  include the $b$-quark contribution to both CC and NC scattering, and are based on  updated PDFs derived from newer and larger data sets.

In Fig.\ \ref{fig:compareplot}  we compare our UHE cross sections from Fig.\ \ref{fig:heraCC-NC} with those of Cooper-Sarkar, Mertsch, and Sarkar  \cite{c-sms},  who used the HERA-based PDF set HERAPDF1.5, and  included the $b$ quark but not the $t$ in their computations.  Their quoted error estimates are in the 2\%-4\% range, comparable to ours, when  they exclude those PDF sets which lead to an unacceptably steep rise in the cross section or allow negative values of the gluon PDF at small $x$ and small $Q^2$.
%
\begin{figure}[htbp]
\includegraphics{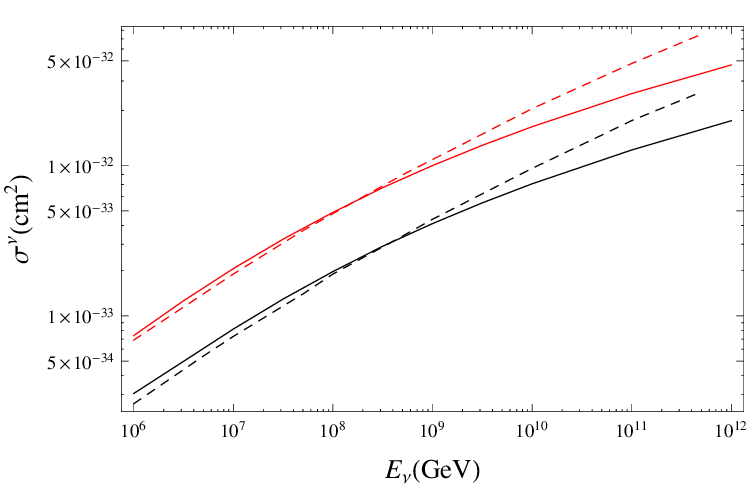}
 \caption{Plots of $\nu N$ cross sections, in cm$^2$, vs. $E_\nu$, the laboratory neutrino energy, in GeV. Our CC cross section is the upper solid (red) curve and our NC cross section is the lower solid (black) curve;  the  CC cross section of  Ref. \cite{c-sms} is the upper dashed (red) curve and their NC cross section is the lower dashed (black) curve. The energy range is that reported in Ref. \cite{c-sms}.  All cross section calculations include the b-quark.}
   \label{fig:compareplot}
\end{figure}
%

The Froissart bound based  and PDF based calculations agree very well for $E_\nu\approx 10^{8}$-$10^{9}$ GeV, where essentially the entire neutrino cross section arises from regions in $x$ (Fig.\ \ref{fig:HERAx-dist}) and $Q^2$ (Fig.\ \ref{fig:HERAQsq-dist}) corresponding to the $x,\,Q^2$ region of the HERA data where the $b$-quark  is reasonably above its excitation threshold and valence-quark contributions to $F_2^{\gamma p}$ are small.

At $E_{\nu}$ = $10^{11}$ GeV and $5\times10^{11}$ GeV, the highest energies reported by Cooper-Sarkar et al.\ \cite{c-sms}, their cross sections  are a factor two larger than ours and extrapolate to more than an order of magnitude larger than ours at $E_{\nu} = 10^{16}$ GeV. This large difference results from their use of a power-law extrapolation in $x$ of the HERAPDF1.5  parton distributions, whereas our partons are bounded by the saturated Froissart-bound  fit to $F_2^{\gamma p}(x,Q^2)$, and grow only as $\ln^2(1/x)$.     A measurement of the CC cross section for $E_\nu\gtrsim{\rm few}\times 10^{11}$ GeV could provide a crucial test of these results.

To provide points of comparison with the other PDF-based calculations mentioned above,  we re-tabulate our results together with those of CSMS, CTW, and GQRS in Tables \ \ref{table:CCx-secs} and \ref{table:NCx-secs}. The tables cover the cross section up to the  $E_{\nu}=10^{12}$ GeV values  published in GQRS and CTW.  Since CSMS quotes cross section values only up to $E_{\nu}=5\times 10^{11}$ GeV, we enter them at  $E_{\nu}=10^{12}$ and note them with an asterisk.  The CSMS, CTW, and GQRS cross sections are all much larger than ours for $E_\nu\gtrsim 10^{-11}$, and would presumably continue to grow much more rapidly than ours at still higher neutrino energies.

\begin{table}[ht]                   
%
\begin{center}
     \caption{Charged current $\nu N$ cross sections, in cm$^2$, as a function of $E_\nu$, the laboratory energy, in GeV:  $\sigma_{BDHM}$ from this work ;   $\sigma_{CTW}$ from \cite{ctw} ;  $\sigma_{CSMS}$ from \cite{c-sms} ; $\sigma_{GQRS}$ from \cite{gandhi98}. Note that CTW and GQRS present values only up to $E_{\nu}=10^{12}$ GeV, while CSMS give values up to $5\times 10^{11}$ GeV.  The entry at $E_{\nu} = 10^{12}$ GeV is marked by an asterisk to signal that the energy for $\sigma_{SMS}$ is actually $5\times 10^{11}$ GeV. \label{table:CCx-secs}
 }
   \renewcommand{\arraystretch}{1.5}
   \begin{tabular}{|c|l|l|l|l|l|r|}
 \hline

      $E_\nu$ (GeV)    & $\sigma_{BDHM}({\rm cm^2})$ & $\sigma_{CTW}({\rm cm^2})$  & $\sigma_{CSMS}({\rm cm^2})$ & $\sigma_{GQRS}({\rm cm^2})$ \\  \hline

           $10^6$    & $7.40\times 10^{-34}$ & $7.2 \times 10^{-34}$ & $6.9\times 10^{-34}$ & $ 6.34\times 10^{-34}$ \\  \hline

            $10^7$    & $2.07\times 10^{-33}$ & $2.0\times 10^{-33}$ & $1.9\times 10^{-33}$ & $1.75\times 10^{-33}$ \\   \hline

            $10^8$    & $4.89\times 10^{-33}$ & $4.8 \times 10^{-33}$ &$ 4.8\times 10^{-33}$ &$ 4.44\times 10^{-33}$ \\   \hline

            $10^9$    & $1.00\times 10^{-32}$ & $1.1\times 10^{-32}$ & $1.1\times 10^{-33}$ & $1.05\times 10^{-32}$ \\   \hline

            $10^{10}$  & $1.82\times 10^{-32}$ & $2.2\times 10^{-32}$ & $2.4\times 10^{-32}$ & $2.38\times 10^{-32}$ \\  \hline

            $10^{11}$   & $3.02\times 10^{-32}$ & $4.3\times 10^{-32}$ & $4.8\times 10^{-32}$ & $5.34\times 10^{-32}$ \\  \hline

            $10^{12}$   & $4.69\times 10^{-32}$ &$ 8.3\times 10^{-32}$ & $7.5\times 10^{-32}$* & $1.18\times 10^{-31}$ \\  \hline

        \end{tabular}
        \renewcommand{\arraystretch}{1}
        \end{center}
\end{table}
%

 \begin{table}[ht]

\begin{center}

    \caption{Neutral current $\nu N$ cross sections, in cm$^2$, as a function of $E_\nu$, the laboratory energy, in GeV: $\sigma_{BDHM}$, from this work;  $\sigma_{CTW}$, from \cite{ctw}; $\sigma_{CSMS}$ from \cite{c-sms}; $\sigma_{GQRS}$ from  \cite{gandhi98}. Note that CTW and GQRS only evaluate values  up to $E_{\nu}=10^{12}$ GeV, while CSMS give values up to $5\times 10^{11}$ GeV.  The entry at $E_{\nu} = 10^{12}$ GeV is marked by an asterisk to signal that the energy for $\sigma_{SMS}$ is actually  $5\times 10^{11}$ GeV. \label{table:NCx-secs}
}
   \renewcommand{\arraystretch}{1.5}
   \begin{tabular}{|c|l|l|l|l|r|}

\hline

      $E_\nu$ (GeV)    & $\sigma_{BDHM}({\rm cm}^2)$ & $\sigma_{CTW}({\rm cm}^2)$& $\sigma_{CSMS}({\rm cm}^2)$ & $\sigma_{GQRS}({\rm cm}^2)$ \\  \hline

           $10^6$    &$ 3.04\times 10^{-34}$ &$ 2.7 \times 10^{-34}$ &$2.6\times 10^{-34}$ &$2.60\times 10^{-34}$ \\   \hline

            $10^7$    &$ 8.17\times 10^{-34}$ &$ 7.6\times 10^{-34}$ &$7.3\times 10^{-34}$ &$7.48\times 10^{-34}$ \\   \hline

            $10^8$    &$ 1.97\times 10^{-33}$ &$ 1.9\times 10^{-33}$ &$1.9\times 10^{-33}$ &$1.94\times 10^{-33}$ \\   \hline

            $10^9$    &$ 4.12\times 10^{-33}$ &$ 4.3\times 10^{-33}$ &$4.4\times 10^{-33}$ &$4.64\times 10^{-33}$ \\   \hline

            $10^{10}$  &$ 7.58\times 10^{-33}$ &$ 9.0\times 10^{-33}$ &$9.6\times 10^{-33}$ &$1.07\times 10^{-32}$ \\  \hline

            $10^{11}$   &$ 1.27\times 10^{-32}$ &$ 1.8\times 10^{-32}$ &$2.0\times 10^{-32}$ &$ 2.38\times 10^{-32}$ \\  \hline

            $10^{12}$   &$ 2.00\times 10^{-32}$ &$ 3.5\times 10^{-32}$ &$3.1\times 10^{-32}$* &$5.20\times 10^{-32}$ \\  \hline

        \end{tabular}
        \renewcommand{\arraystretch}{1}
      \end{center}
\end{table}

As already noted, CSMS use PDFs from a fit to the combined HERA results, so the data used and the inclusion of the $b$-quark make their work the most natural to compare to ours. CTW use the  MSTW2008 NLO PDFs, which bases its small-$x$ information on ZEUS data and provides PDF grids down to $x$ = $10^{-6}$;  for smaller $x$, CTW extrapolate the quark PDFs with a form $a+b\ln(1/x)$, $a$ and $b$ constants, rather than  the power-law form that follows from the extension of the initial MSTW quark parametrizations to small $x$. Use of the latter would  lead to much larger cross sections than CTW obtained.

All these PDF calculations use extrapolations of individual quark distributions to values  of $x$ well below the region of the HERA data using somewhat arbitrary assumptions about their $x$ dependence. Our extrapolation, in contrast, is of the  structure function $F_2^{\gamma p}$ -- that is, the virtual-boson, hadron scattering cross section --  using the Froissart-bounded form which is favored by theory and provides an excellent fits not only to the $\gamma^*p$ HERA data, but to the $\gamma p$ and hadronic data to the highest energies studied. We regard this extrapolation, by a factor of $\sim 2.7$ in the natural variable $v=\ln(1/x)$, as far more reliable theoretically and numerically. Our quark distributions follow from this fit.  We conclude that the cross sections predicted by the PDF-based calculations with power-law extrapolations to  small $x$ are unrealistically large at ultra-high energies.


\section{Implications for hadronic physics \label{sec:hadronic_implications}}


\subsection{Probing ultra-high hadronic energies \label{subsec:probingUHE}}


We remark finally on the implications of these results for hadronic physics. Our fundamental assumption, discussed in detail in Part I \cite{bdhmFroissart}, is that the virtual gauge boson--hadron scattering processes $\gamma^*p$, $W^*N$ and  $Z^*N$ are basically hadronic in nature, each interaction having the same Froissart-bounded structure seen in all very high energy hadronic cross sections, including real $\gamma p$ scattering \cite{mbair,blockhalzen,blockhalzen2}. The structure functions $F_2^{\gamma p}$, $F_2^{\nu(\bar{\nu})}$, and $F0_2^{\nu(\bar{\nu})}$ for $ep$ and CC and NC $\nu(\bar{\nu}) N$ scattering for a virtual boson mass $-Q^2$ are all related, differing only in the electromagnetic and weak charges  and the helicity structure of  the quark currents.   We therefore expect the {\em same} Froissart-bounded structure in neutrino as in electron interactions. However, the final hadronic cms energies $W=\sqrt{\hat{s}}$ potentially accessible and measurable in UHE neutrino interactions range far above the energies which have been studied in other experiments, 7 TeV at the Large Hadron Collider \cite{LHCtot1,LHCtot2,LHCtot3} and 57 TeV at the Pierre Auger cosmic ray array \cite{POAp-air}. This is shown in Fig.\ \ref{fig:Waverage_vs_E_nu}, where we plot both $W_{\rm rms}$, the square root of the average of $\hat{s}=Q^2/x$, and the average value of the final cms hadronic energy  $\hat{W}=\sqrt{\hat{s}}$, versus the incident neutrino energy, with the averages taken over the $F_2^{\nu(\bar{\nu})}$ contribution to the CC neutrino cross section.
%
\begin{figure}[htbp]
\includegraphics{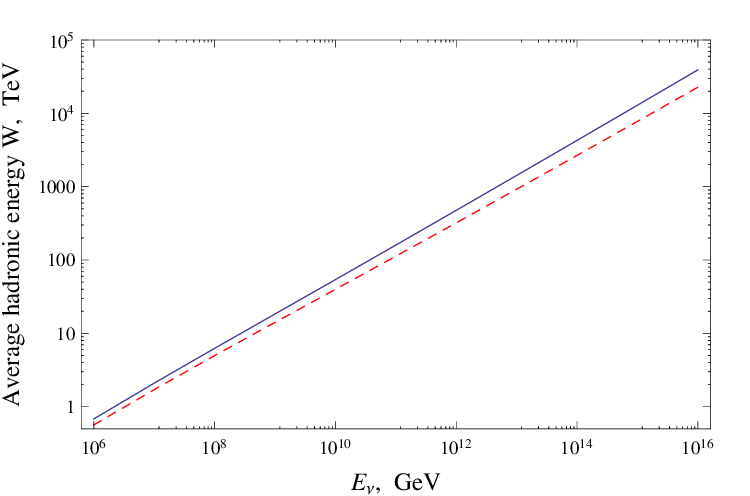}
\caption{Plots of average final hadronic cms energies explored in UHE $\nu N$ scattering.  $W_{av}=(\sqrt{\hat{s}})_{av}$ (dashed red curve), and  $W_{\rm rms}=\sqrt{\hat{s}_{av}}$ (solid blue curve), in TeV, are plotted as functions of the incident neutrino energy $E_{\nu}$, in GeV. Here $\hat{s}=Q^2/x$. The averages are over the charged current ($Q^2$, $x$) distribution in \eq{masterEq2cc}. Note the enormously high hadronic cms energies that are available, over  $10^4$ TeV for $E_{\nu}\sim 10^{16}$ GeV. }

   \label{fig:Waverage_vs_E_nu}
\end{figure}
%
It is evident from this log-log plot that $W_{av}$ increases nearly as a power with $E_\nu$, approximately as $E_\nu^{0.46}$, close to  the behavior $W_{max}=\sqrt{2mE_\nu}$ attained at the upper kinematic limit.

The average cms hadronic energy $W_{av}$ in $\nu N$ scattering at $E_\nu=10^{11}$ GeV is 113 TeV, already twice the 57 TeV reached by the Auger Collaboration, while at $E_\nu=10^{16}$ GeV, $W_{av}=\ $22,675 TeV. As seen from Fig.\ \ref{fig:compareplot} and Tables\ \ref{table:CCx-secs} and \ref{table:NCx-secs}, the effects of the Froissart bound are clearly evident at these energies, with our predicted cross sections substantially lower than those obtained in standard approaches based on the DGLAP evolution of quark PDFs, extrapolated to small $x$. The striking differences provide a clear test of the underlying ideas.


\subsection{Analytic form of the CC and NC cross sections as functions of $E_\nu$\label{analytic}}


In Fig. \ref{fig:logcube}
we plot the CC (upper points) and NC (lower points) cross sections of Table \ref{tab:errors}, in mb, vs. $E_\nu$ in GeV. We fit a 4 parameter fit, of the form  $\sigma= a+b\ln E_\nu+c \ln E_\nu^2+d\ln E_\nu^3$, to the 12 points of Table \ref{tab:errors} to obtain the analytic cross section functions
\ba
\sigma_{CC}(E_\nu)&=&-19.91+4.685\ln E_{\nu}-0.3798\  \ln^2 E_\nu+0.01078\  \ln^3 E_\nu,\label{sigCCanalytic}\\
\sigma_{NC}(E_\nu)&=&-10.161+2.304\ln E_{\nu}-0.1801 \  \ln^2 E_\nu+0.004926\  \ln^3 E_\nu,\label{sigNCanalytic}
\ea
with $E_\nu$ in GeV and the constants and cross sections in mb.
\begin{figure}[htbp]
\includegraphics{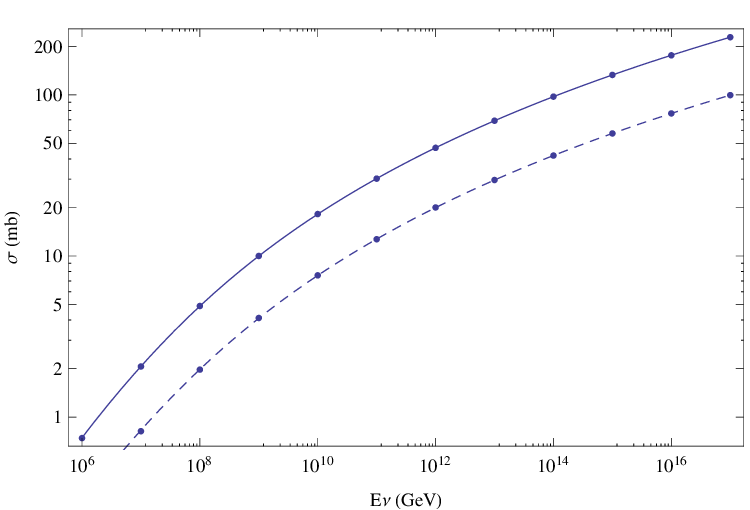}
 \caption{Plots of $\nu N$ cross sections, in cm$^2$, vs. $E_\nu$, the laboratory neutrino energy, in GeV.  Our analytic CC cross section  is the upper solid  curve of \eq{sigCCanalytic} and the analytic NC cross section is the lower dashed curve of \eq{sigNCanalytic}. The points are the numerical calculations of Table \ref{tab:errors}.  The agreement is better than 1 part in 1000.
 }
   \label{fig:logcube}
\end{figure}

The upper solid curve in Fig.\ \ref{fig:logcube} is the $\ln^3E_\nu$ parameterization of $\sigma_{CC}$ and the lower dashed curve is  the $\ln^3E_\nu$ parameterization of $\sigma_{NC}$.  The points are the numerical calculations of Table \ref{tab:errors}.
Clearly, the $\ln^3 E_\nu$  parameterization is excellent, with numerical agreement better than 1 part in 1000.

A  discussion of why a strong interaction Froissart bound of $\ln^2 (1/x)$ gives rise to a weak interaction $\nu N$ cross section bound of $\ln^3 E_\nu$ is given in the last paragraphs of   Section  \ref{subsec:total_sigmas} and in \cite{kniehl}. Conversely, a weak cross section $\nu N$ bound of $\ln^3 E_\nu$ {\em  implies} a strong cross section hadron-nucleon Froissart bound of $\ln^2 \hat s=\ln^2 W^2$.

An experimental demonstration that $\nu N$ cross section measurements in this energy region are bounded by $\ln^3 E_\nu$ would give a clear test of the entire picture discussed here. It would show that   the virtual boson-hadron interactions are hadronic in nature,  having the same Froissart-bound saturating structure as observed in other hadronic interactions, and thus allow  future experiments to use weak $\nu N$ interactions to explore strong hadronic interactions at otherwise unobtainable energies.


\section{Summary and conclusions \label{conclusions}}


In response to the ongoing need for theoretical calculations to guide design of ultra-high energy neutrino telescopes, we have improved our earlier calculations of the UHE $\nu N$ total cross section given in \cite{bhm}.  The results include the b-quark contribution, whose importance we have shown in our analysis of the kinematical region that dominates the total cross section integral. The results are based on the relation between the dominant neutrino structure function $F_2^{\nu(\bar{\nu})}$ and the $\gamma p$ structure function $F_2^{\gamma p}$ established in Part I \cite{bdhmFroissart},  and corresponding relations for the subdominant structure functions $F_3^{\nu(\bar{\nu})}$ and $F_L^{\nu(\bar{\nu})}$  correct to NLO.

The basic assumption, justified in Part I, is that  $F_2^{\gamma p}$,  the  $\gamma^*p$ reduced cross section for the interaction of an  off-shell photon  with a proton, has the Froissart-bounded form found in hadronic scattering and $\gamma p$ scattering. This provides an excellent fit to the HERA data on $ep$ DIS, even  including the regions where $Q^2$ is so small that pQCD is not expected to be valid. The Froissart form can be extrapolated reliably through the factor $\sim2.7$ extension of the range of the natural variable $\ln(1/x)$ needed to reach the $x$ values appropriate to UHE neutrino scattering up to $E_\nu=10^{17}$ GeV.

The uncertainties in our results arise primarily from the uncertainties in the values of the parameters in our saturated  Froissart-bounded fit  to the combined HERA data \cite{HERAcombined} on $F_2^{\gamma p}(x,Q^2)$; they amount to 1\% - 2\% uncertainty in our cross section values over the entire $E_{\nu}$ range  that we study. We show that our $\nu N$ cross sections are bounded by $\ln^3 E_\nu$, which is a {\em consequence} of our Froissart-bounded  $\ln^2(1/ x)$ fit to $F_2^{\gamma p}(x,Q^2)$.

Comparing our cross section values to those of the most recent PDF-based study \cite{c-sms}---which also includes the effects of the b-quark---we find that our cross sections are significantly lower than theirs at high neutrino energies, with the results diverging strongly for neutrino energies above $\sim 10^{11}$ GeV.

In conclusion, we believe the neutrino cross sections calculated starting from   our saturated Froissart bound  fit to the existing experimental HERA data are the most physically motivated, and  thus provide the best estimates of  UHE energy $\nu N$ cross sections now  possible. Moreover,  these UHE neutrino interactions have significant potential implications for hadronic physics up to an average ultra-high hadronic cms energy  $W=\sqrt {\hat s}\gtrsim 70,000$ TeV, if we can discover and measure $\nu N$  cross sections  with laboratory energies $E_\nu\sim 10^{17}$ GeV.


\begin{acknowledgments}

 M.\ M.\ B.\ and L.\ D.\ would like to thank the Aspen Center for Physics, where this work was  supported in part by NSF Grant No. 1066293, for its hospitality.
 M.\ M.\ B.\  would like to thank Profs.\ A.\ Vainshtein and G.\ Domokos for valuable discussions.   P.\ H.\ would like to thank Towson University Fisher College of Science and Mathematics for support.  D.\ W.\ M.\ received support from DOE Grant No. DE-FG02-04ER41308.

\end{acknowledgments}


\appendix

\section{The dominant structure functions $F_2^{\nu(\bar{\nu})}$ at NLO  \label{sec:F2}}


 We recall that the result for the dominant structure function $F_2^{\nu(\bar{\nu})}$ for different numbers of active quarks  derived in Part I, is given by
\ba
\label{F2nu_F2_connection_nf=3}
F_2^{\nu(\bar{\nu})}  &=& \frac{9}{2}F_{20}^{\gamma p}-\frac{1}{4}T'_8-\frac{3}{8}U',  \qquad\qquad\qquad  \qquad  \qquad n_f=3,\quad Q_0^2< Q^2\leq m_c^2, \\
\label{F2nu_F2_connection_nf=4}
F_2^{\nu(\bar{\nu})}  &=&  \frac{18}{5}F_{20}^{\gamma p} -\frac{1}{5}T'_8+\frac{1}{5}T'_{15}-\frac{3}{10}U',\quad\qquad  \qquad  \quad n_f=4,\quad m_c^2< Q^2\leq m_b^2,  \\
\label{F2nu_F2_connection_nf=5}
F_2^{\nu(\bar{\nu})}  &=& \frac{45}{11}F_{20}^{\gamma p} -\frac{5}{22}T'_8+\frac{5}{22}T'_{15}-\frac{3}{22}T'_{24} -\frac{15}{44}U', \quad n_f=5,\quad m_b^2< Q^2.
\ea
Here $U$ is the valence quark distribution in the approximation $d=(1/2)u$ and the $T$s are the non-singlet combinations of quark distributions defined, for example, in Ref.  \cite{esw}, and used in Part I. The primes indicate that $U$ and the $T'$s have been transformed from the quark level to that of the observable structure functions using the NLO corrections from the operator product expansion in Refs. \cite{hw,furmanski,esw}.

The manipulations which connect  $F_2^{\nu(\bar{\nu})}$ to $F_2^{\gamma p}$, culminating in Eqs.\ (\ref{F2nu_F2_connection_nf=3})--(\ref{F2nu_F2_connection_nf=5}), can be handled analytically at small $x$ as discussed in detail in the Appendices in Part I \cite{bdhmFroissart}. The final results for the leading $F_2^{\gamma p}$ terms are  correct formally to all orders in $\alpha_s$. The theoretical uncertainties in the final result arise through the non-singlet functions $U'$, $\hat{T}'_8$, $\hat{T}'_{15}$ and $\hat{T}'_{24}$, which we have only treated to NLO.

In our calculations, the valence distribution $U$ was taken from CTEQ5 \cite{cteq5} and the (small) function $T_8$ was taken from the result from the  CT10 analysis of the HERA and other data \cite{CT10}, extrapolated to small $x$ using the Froissart bound form of the fit function required by consistency with the form of $F_2^{\gamma p}$. They were used as an input to get the singlet distribution $F_s = \sum_i(q_i+\bar{q}_1)$ from $F_{20}^{\gamma p}$, known to NLO, at $Q^2=m_c^2$. This determines $T_{15}(x,m_c^2)=F_s(x,m_c^2)$. As shown in Part I, the changes in $T_8$ and $T_{15}$ induced by QCD evolution to $Q^2=m_b^2$ are minimal and can be calculated analytically to sufficient accuracy for our purposes; $T_{24}(x,m_b^2)$ is then determined from $F_{20}^{\gamma p}$, $T_8$, and $T_{15}$. The evolution of $T_{24}$ is again known analytically.

There is a further uncertainty in the $T_i$ in that  the transformation from $F_2^{\gamma p}$ to the uncorrected distribution $F_{20}^{\gamma p}$ needed in the determination of $F_s$ depends on the gluon distribution $g$, which we again took from the CT10 analysis as extrapolated to small $x$. We estimate the overall uncertainties in the small non-singlet $T$ terms in  Eqs.\ (\ref{F2nu_F2_connection_nf=3})--(\ref{F2nu_F2_connection_nf=5}) to be significantly less than the $\lesssim 10$\% difference between $F_2^{\gamma p}$ and our derived $F_{20}^{\gamma p}$, and to lead to  at most a 1-2\% uncertainty in the final results for $F_2^{\nu(\bar{\nu})}$.

The analog  of \eq{F2nu_F2_connection_nf=5} for $F0_2^{\nu(\bar{\nu})} $for $Q^2>m_b^2$ and $n_f=5$ is
\ba
F0_2^{\nu(\bar{\nu})} &=& \frac{9}{22}\left(F_2^{\gamma p}-\frac{1}{12}U'\right)\left[3\left(L_d^2+R_d^2\right)+2\left(L_u^2+R_u^2\right)\right]  \nonumber \\
\label{F02nu_F2_connection_smallx}
&& -\frac{1}{132}\left[4\left(L_d^2+R_d^2\right)-\left(L_u^2+R_u^2\right)\right]\left(5T'_8-5T'_{15}+3T'_{24}\right).
\ea
The contributions associated with the $T'$s are negative and decrease the final results for $F0_2^{\nu(\bar{\nu})}$ by
 $\sim$5.9\% to 4.3\% as $x$ decreases from $10^{-5}$ to $10^{-14}$ at $Q^2=100$ GeV$^2$, and by 1.8\% to 1.1\% over the same $x$ range for $Q^2=10,000$ GeV$^2$.

It is interesting to note the connection of these results to  Feynman's  wee parton picture as applied to neutrino interactions. In this picture, the quark distributions all converge toward a common distribution at small $x$ where the valence quark contributions are negligible  and sea quark distributions are all  equal (equipartition of flavors), $xq_i(x,Q^2)\rightarrow xq(x,Q^2)$ for all $i$. The $T'$s  and $U'$ then vanish individually at small $x$, and $F_2^{\nu(\bar{\nu})}$ reduces to a multiple of $F_2^{\gamma p}$,
\be
\label{CCwee}
F_2^{\nu(\bar{\nu})} \approx \left[\left(\sum_ic_{2,i}\right)\Big/\left(\sum_ie_i^2\right)\right]F_2^{\gamma p},
\ee
for the charged current interactions, with a similar result for neutral current interactions.

This approximation was used in Refs.\  \cite{bbmt, bhm}, where it was established only in LO where $xq(x,Q^2)=F_2^{\gamma p}(x,Q^2)/\sum_ie_i^2$,  to predict UHE neutrino cross sections for $n_f=4$ in terms of $F_2^{\gamma p}$, neglecting potential QCD corrections, the contribution of the $b$ quark and the small terms associated with the structure functions  $F_3^{\nu(\bar{\nu})}$ and  $F_L^{\nu(\bar{\nu})}$.  We compare the predictions of the (supposed) wee parton model  for $n_f=5$, now including the contribution of  $F_3^{\nu(\bar{\nu})}$ in the wee limit $F_3^{\nu(\bar{\nu})}\approx (9/11)F_2^{\gamma p}$, with the complete results from the present calculation in Table \ref{table:exact_wee_comp}.

 While these simplified results are strikingly good and accurate enough for most purposes, we showed in Part I that the wee parton limit actually does not exist: it is upset by the different thresholds at  $Q_i^2=m_i^2$ at which the various quarks become active. The quark distributions for the $s$, $c$, and $b$ quarks actually diverge from each other and from the light-quark distributions as $\ln^2(1/x)$ with decreasing $x$, and the $T$s are nonzero.  However,  the combination of $T'$s in Eqs.\ (\ref{F2nu_F2_connection_nf=3})-(\ref{F2nu_F2_connection_nf=5}) turns out to be quite small relative to $F_{20}^{\gamma p}$  because of cancellations, and the effective wee parton approximation in \eq{CCwee} is still useful.

 We emphasize, however, that it is not difficult to use the results on the quark distribution from Part I to evaluate the cross sections in full. Furthermore, it is {\em essential} to establish that the corrections to the wee parton relations are small before it is used in a different context.

%
\begin{table}[htbp]
   \begin{center}
        \renewcommand{\arraystretch}{1.5}
    \begin{tabular}{|l||c|c|c|c|c|c|}
         \hline
       Energy  (GeV) &$10^6$ &$10^8$ &$10^{10}$ & $10^{12}$ & $10^{14}$ & $10^{16}$  \\ \hline\hline
        $\sigma_{\rm CC,\ exact},{\rm \ cm}^2$ &$7.40\times 10^{-34}$ &$4.89\times 10^{-33}$ &$1.82\times 10^{-32}$ &$4.69\times 10^{-32}$  &$9.75\times 10^{-32}$ &$1.76\times 10^{-31}$ \\ \hline
        $\sigma_{\rm CC,\ wee},{\rm \ cm}^2$ &$7.54\times 10^{-34}$ &$5.00\times 10^{-33}$ &$1.84\times 10^{-32}$ &$4.74\times 10^{-32}$  &$9.84\times 10^{-32}$ &$1.78\times 10^{-31}$ \\ \hline
        $\sigma_{\rm NC, \ exact},{\rm \ cm}^2$&$3.04\times 10^{-34}$ &$1.97\times 10^{-33}$ &$7.58\times 10^{-33}$ &$2.00\times 10^{-32}$  &$4.20\times 10^{-32}$ &$7.67\times 10^{-32}$ \\ \hline
        $\sigma_{\rm NC,\ wee},{\rm \  cm}^2$ &$3.12\times 10^{-34}$ &$2.02\times 10^{-33}$ &$7.70\times 10^{-33}$ &$2.02\times 10^{-32}$  &$4.24\times 10^{-32}$ &$7.73\times 10^{-32}$ \\ \hline
    \end{tabular}
 \renewcommand{\arraystretch}{1}

   \end{center}
   \caption{Comparison of the exact CC and NC cross sections for $n_f=5$ with their wee parton approximations.}
   \label{table:exact_wee_comp}
\end{table}


\section{The sub-dominant structure functions $F_3^{\nu(\bar{\nu})}$ and $F_L^{\nu(\bar{\nu})}$ at NLO  \label{subsec:F3FL}}


Although the $F_2^{\nu(\bar{\nu})}$ structure function is dominant in  $\nu N$ and $\bar\nu$-$N$ UHE scattering as was discussed in Sec.\ \ref{subsec:sensitivity}, the contribution of the structure function $F_3^{\nu(\bar{\nu})}$ through the $b$ quark is significant;  we wish to estimate it at NLO, and further, to include the NLO contribution of $F_L^{\nu(\bar{\nu})}$, which is zero at LO in $\alpha_s$.  We will concentrate on the region $Q^2 > m_b^2$, which contributes all but a small fraction of the cross sections, as we saw in Sec.\ \ref{ultrahighsigma}.

We start by re-expressing $xF_{30}^{\nu(\bar{\nu})}$, given at the quark level in \eq{F3}, in terms of $F_{20}^{\nu(\bar{\nu})}$, $U$, and the $T_i$:
\ba
\label{F3_F20relation3}
xF_{30}^{\nu(\bar{\nu})} &=& \frac{3}{2}F_{20}^{\gamma p} - \frac{5}{12}T_8+\frac{11}{8}U,\qquad n_f=3, \\
\label{F3_F20relation4}
xF_{30}^{\nu(\bar{\nu})} &=& \frac{1}{3}\left(T_{15}-T_8\right)+\frac{3}{2}U,\qquad n_f=4, \\
\label{F3_F20relation5}
xF_{30}^{\nu(\bar{\nu})} &=& \frac{9}{11}F_{20}^{\gamma p}  - \frac{5}{66}\left(5T_8-5T_{15}+3T_{24}\right)+\frac{63}{44}U,\qquad n_f=5.
\ea
The $T$ terms in the expression for $xF_{30}^{\nu(\bar{\nu})}$ for $n_f=5$ appear in the same combination as in the expression for $F_{2}^{\nu(\bar{\nu})}$ in \eq{F2nu_F2_connection_nf=5}, but with a coefficient which is larger relative to the coefficient of $F_{20}^{\gamma p}$ by a factor 25/3, and a valence term of the opposite sign. As a result, the combination of $T_i$ and $U$, previously quite small relative to $F_{20}^{\gamma p}$, is now significant and should be taken into account.

To obtain the physical structure function $F_3^{\nu(\bar{\nu})}$, we must convolute  $F_{3,0}^{\nu ,\bar \nu}$  with the QCD coefficient function $C_{3q}$ \cite{hw,furmanski,esw},
\ba
 \label{NLOxF3}
F_3^{\nu(\bar{\nu})}&=& F_{3,0}^{\nu(\bar{\nu})}+\frac{\alpha_s}{ 2 \pi}C_{3q}\otimes F_{3,0}^{\nu(\bar{\nu})} .
\ea
The  gluon does not enter and no assumption about $g$ is necessary even though $F_{3,0}^{\nu(\bar{\nu})}$ has a mixed singlet, non-singlet structure for $b=\bar{b}$ nonzero \cite{bardeen}. Since the gluon does not appear and $C_{3q}\not=C_{2q}$, the transformation does not convert $F_{20}^{\gamma p}$ to $F_2^{\gamma p}$ as in the calculations above. However, the convolution integrals are readily evaluated at small $x$ using the methods of Part I \cite{bdhmFroissart}  once $xF_{3,0}^{\nu(\bar{\nu})}$ is known; see Appendix  B of Part I.

 The contribution of $xF_3^{\nu(\bar{\nu})}$ to the cross sections is described in Sec. \ref{subsec:sensitivity}.  We note here that at small $x$, where the valence term $U$ is negligible,  $xF_3^{\nu}(x, Q^2)=-xF_3^{\bar \nu}(x, Q^2)$. As a result,  $\sigma_{CC}^{\nu}\approx \sigma_{CC}^{\bar \nu}$ for large $E_\nu$  despite the presence of the $\pm$ signs in \eq{masterEq2cc}.

 The NC structure function  $F0_{30}^{\nu(\bar{\nu})}$, \eq{F30NC}, depends only on the valence quark distributions $u_v=u-\bar{u}=U$ and $d_v=d-\bar{d}\approx U/2$, so vanishes at small $x$ where $u\sim\bar{u}$ and $d\sim\bar{d}$, and does not contribute to the NC cross section there.

The structure function $F_{L}^{\nu(\bar{\nu})}(x,Q^2)$, which is zero in LO, is given in NLO in Ref. \cite{esw}  by
\be
 \label{FLNLO}
x^{-1}F_{L}^{\nu(\bar{\nu})}(x,Q^2) = \frac{\alpha_s}{2\pi} C_{Lq}\otimes(x^{-1}F_{20}^{\nu(\bar{\nu})})+\frac{\alpha_s}{2\pi}2n_f\,C_{Lg}\otimes g. \\
\ee
 The integrals can again be calculated analytically for small $x$ following the methods outlined in the Appendix B to Ref.  \cite{bdhmFroissart}. It is sufficient for our purposes to approximate $F_{20}^{\nu(\bar{\nu})}$ in this calculation by  $(45/11)F_{20}^{\gamma p}$ for $n_f=5$, as in \eq{CCwee}.   We include  $F_{L}^{\nu(\bar{\nu})}$ in the cross section calculations in Sec. \ref{subsec:predictions}.

 A similar result holds for the NC structure function $F0_{L}^{\nu(\bar{\nu})}(x,Q^2)$, with  $F_{20}^{\nu(\bar{\nu})}$ replaced in \eq{FLNLO} by $F0_{20}^{\nu(\bar{\nu})}\approx (9/11)F_2^{\gamma p}\left[L_u^2+R_u^2+(3/2)(L_d^2+R_d^2)\right]$.


\bibliography{small_x_references}

\end{document}